\documentstyle[prd,aps]{revtex}
\tighten
\draft
\begin{document} 
\bibliographystyle{prsty}

\def\Journal#1#2#3#4{{#1} {\bf #2} (#4) #3}
\def\MPL{Mod. Phys. Lett. A}
\def\NCA{Nuovo Cimento}
\def\NPB{Nucl. Phys. B}
\def\NPBOLD{Nucl. Phys.}
\def\NPSUPPL{Nucl. Phys. Proc. Suppl.}
\def\PLB{{Phys. Lett.} B}
\def\PREPO{Phys. Rep.}
\def\PLBOLD{Phys. Lett.}
\def\PRL{Phys. Rev. Lett.}
\def\RMP{Rev. Mod. Phys.}
\def\PRD{Phys. Rev. D}
\def\ZPC{Z. Phys. C}
\def\PTPSUPPL{Prog. Theor. Phys. Suppl.}
\def\PTP{Prog. Theor. Phys.}
\def\JHEP{JHEP}
\def\EPJ{Euro. Phys. J. C}
\def\JETPUSSR{JETP (USSR)}
\def\ZETP{Zh. Eksp. Teor. Piz.}
\def\TNYAS{Trans. New York Acad. Sci.}
\def\IJMP{Int. J. Mod. Phys. A}
\def\PPNP{Prog. Part. Nucl. Phys.}
\def\ARNPS{Ann. Rev. Nucl. Part. Sci.}

\mathchardef\ddash="705C

\title{
Physics of Strongly Coupled $N$=1 Supersymmetric $SO(N_c)$ Gauge Theories
}

\author{Masaki Yasu\`{e}
\footnote{E-mail:yasue@keyaki.cc.u-tokai.ac.jp}
}

\address{\vspace{2mm}{\sl General Education Program Center, 
 Shimizu Campus, 
\\School of Marine Science and Technology, Tokai University}\\
{\sl 3-20-1 Orido, Shimizu, Shizuoka 424-8610, Japan}}
\address{{\sl and}\\{\sl Department of Physics, Tokai University}\\
{\sl 1117 KitaKaname, Hiratsuka, Kanagawa 259-1292, Japan}}
\maketitle

\begin{abstract}
We construct an effective superpotential that describes dynamical flavor symmetry breaking in supersymmetric $N$=1 $SO(N_c)$ theories with $N_f$-flavor quarks for $N_f \geq N_c$. Our superpotential induces spontaneous flavor symmetry breaking of $SU(N_f)$ down to $SO(N_c)$ $\times$ $SU(N_f-N_c)$ as nonabelian residual groups and respects the anomaly-matching property owing to the appearance of massless composite Nambu-Goldstone superfields.  In massive $SO(N_c)$ theories, our superpotential provides holomorphic decoupling property and consistent vacuum structure with instanton effect if $N_c-2$ quarks remain massless.  This superpotential may reflect the remnant of physics corresponding to $N$=2 $SO(N_c)$ theories near the Chebyshev point, which also exhibits dynamical flavor symmetry breaking.
\end{abstract}


The phase structure of strongly coupled $N$=1 supersymmetric (SUSY) gauge theories can be determined by perturbing well-understood $N=2$ gauge theories\cite{SeibergWitten,BrokenN_2}. The $N$=1 supersymmetric quantum chromodynamics (SQCD) with $N_c$-colors and $N_f$-flavors is well described by the $N$=1 duality\cite{Seiberg,EarlySeiberg} for $N_f\geq N_c+2$ using dual quarks.  The similar dual description is also applicable to other gauge theories such as those based on $SO(N_c)$\cite{SO,RecentSO}.  However, it is only possible to theoretically confirm the validity of the $N$=1 duality by embedding it in softly broken $N$=2 theories. Since the $N$=2 theories are so constrained, it is not possible to cover the confining physics for all ranges of $N_f$ for a given $N_c$.  In SQCD with $N_f<3N_c$, the $N$=1 duality is believed to describe its physics for $N_f$ $\geq$ $N_c$+2 as long as dual quarks have $N_f-N_c$ colors\cite{Seiberg,Review} while the $N$=2 duality supports the $N$=1 duality only if $N_f>3N_c/2$, where SQCD is characterized by an interacting Coulomb phase\cite{CoulombPhase}.  

In $SO(N_c)$ gauge theories (with $N_f< 3(N_c-2)$), the similar question on the validity of $N$=1 duality has lately been raised\cite{RecentSO} by noticing the existence of the Chebyshev point in $N$=2 gauge theories.  Physics near the Chebyshev point is found to be characterized by the presence of dynamical flavor symmetry breaking due to the condensation of mesons, which should be contrasted with the absence of flavor symmetry breaking in the $N$=1 duality.  Although, in the $N$=1 limit, the Chebyshev point appears to merge with the ordinary singular point, where the flavor symmetries remain unbroken, we speculate that there is a strictly stable phase with dynamical flavor symmetry breaking even in the $N$=1 limit, which reflects the $N$=2 physics near the Chebyshev point.  

In this paper, we suggest that the low-energy degree of freedom in confining $SO(N_c)$ theories with $N_f \geq N_c$ is provided by Nambu-Goldstone superfields associated with the dynamical flavor symmetry breaking, which is well regulated by our proposed effective superpotential. Although there is no direct proof that out superpotential is related to the $N$=1 physics derived from the $N$=2 physics near the Chebyshev point, we expect that it describes the $N$=1 physics since it passes three consistency checks offered by the holomorphic decoupling property, the instanton physics and the anomaly-matching property\cite{tHooft,AnomalyMatch}.  There have also been several discussions on the possible existence of such a flavor symmetry breaking phase in the case of SQCD with $N_f$ $\geq$ $N_c$+2\cite{Yasue,S,GapEquation,Instability}.

We start with the classic construction of an effective superpotential\cite{Yasue,VY,VYF}, which is invariant under the transformations under all the symmetries and is compatible with the response from an anomalous $U(1)_{anom}$ symmetry.  Namely, we require that not only it is invariant under an anomaly free $SU(N_f)$ $\times$ $U(1)_R$ symmetry, where quark superfields have $(N_f-N_c+2)/N_f$ as a $U(1)_R$ charge, but also it is equipped with the transformation property under $U(1)_{anom}$ broken by the instanton effect, which is represented by $\delta{\cal L}$ $\sim$ $F^{\mu\nu}{\tilde F}_{\mu\nu}$, where ${\cal L}$ represents the lagrangian of the $SO(N_c)$ theory and $F^{\mu\nu}$ (${\tilde F}_{\mu\nu}\sim \epsilon_{\mu\nu\rho\sigma}F^{\rho\sigma}$) is a gauge field strength.   

In the rest of discussions, we follows the same strategy as the one developed in Ref.\cite{Yasue} to examine vacuum structure in the $SO(N_c)$ theory, where we use the freedom that cannot be determined by symmetries and holomorphy\cite{VYF,ISF}.  In $SO(N_c)$ theories with $N_f \geq N_c$, the resulting superpotential, $W_{\rm eff}$, with quark masses included is given by\cite{Honda}
\begin{equation}\label{Eq:Weff}
W_{\rm eff}=S 
\left\{ 
\ln\left[
\frac{
	S^{N_c-N_f+2}{\rm det}\left(T\right) f(Z)
}
{
	\Lambda^{3N_c-N_f-6}
} 
\right] 
+N_f-N_c+2\right\} - \sum_{i=1}^{N_f}m_iT^{ii}
\end{equation}
with an arbitrary function, $f(Z)$, to be determined, where $\Lambda$ is the scale of the $SO(N_c)$ theory and $m_i$ is the mass of the $i$-th quark.  The fields, $S$ and $T$, are composite superfields defined by
\begin{equation}\label{Eq:FieldContentST}
S = \frac{1}{32\pi^2}\sum_{A,B=1}^{N_c} W_{[AB]}W_{[AB]}, \
 T^{ij}  =  \sum_{A=1}^{N_c} Q_A^iQ_A^j,  
\end{equation}
where chiral quark superfields and chiral gauge superfields are, respectively,  denoted by $Q_A^i$ and  $W_{[AB]}$ for $i,j$ = 1 $\sim$ $N_f$ and $A,B$ = 1 $\sim$ $N_c$.  The remaining field denoted by $Z$ describes an effective field defined by
\begin{eqnarray}\label{Eq:FieldContentZB}
Z &=& \frac{\sum_{i_1\cdots i_{N_f}, j_1\cdots j_{N_f}} \varepsilon_{i_1\cdots i_{N_f}}\varepsilon_{j_1\cdots j_{N_f}}B^{[i_1\cdots i_{N_c}]}T^{i_{N_c+1}j_{N_c+1}}\cdots T^{i_{N_f}j_{N_f}}{B}^{[j_1\cdots j_{N_c}]}}
{N_c!N_c!\left(N_f-N_c\right)!\left(N_f-N_c\right)!{\rm det}\left(T\right)},
\end{eqnarray}
where
\begin{eqnarray}\label{Eq:FieldContent}
&&
B^{[i_1i_2{\dots}i_{N_c}]} = \sum_{A_1\dots A_{N_c}}\frac{1}{N_c!}
                                                 \varepsilon^{A_1A_2{\dots}A_{N_c}}Q_{A_1}^{i_1}\dots Q_{A_{N_c}}^{i_{N_c}},
\end{eqnarray}
since we are restricted to study the case with $N_f \geq N_c$.  We use the notation of $Z=BT^{N_f-N_c}B/{\rm det}(T)$. 
Note that $Z$ is neutral under the entire chiral symmetries including $U(1)_{anom}$ and the $Z$-dependence of $f(Z)$ cannot be determined by the symmetry principle.

Two comments are in order: 
\begin{enumerate}
\item Since the $SO(N_c)$ theory has color-antisymmetric gauge fields, there are lots of composite fields containing both quarks and chiral gauge fields.  Whether these composites are light or not can only be examined by the anomaly-matching property of our superpotential normally dictated by the Nambu-Goldstone superfields.  It will be found that these composites need not enter in the low-energy degree of freedom.
\item The explicit use of the redundant field of $S$ is essential in our discussions.  The vacuum structure of the $SO(N_c)$ theory can be more transparent if we study dynamics of the $SO(N_c)$ theory in terms of $S$ by taking a softly broken $SO(N_c)$ theory in its SUSY limit.
\end{enumerate}

It is readily understood that if one flavor becomes heavy, our superpotential exhibits a holomorphic decoupling property. Let the $N_f$-th flavor be massive and others be massless, then we have
\begin{equation}\label{Eq:MassiveWeff}
W_{\rm eff}=S 
\left\{ 
\ln\left[
\frac{
	S^{N_c-N_f-2}{\rm det}\left(T\right) f(Z)
}
{
	\Lambda^{3N_c-N_f-6}
} 
\right] 
+N_f-N_c+2\right\} - mT^{N_fN_f}.
\end{equation}
The field, $T$, can be divided into ${\tilde T}$ with a light flavor $(N_f-1)$ $\times$ $(N_f-1)$ submatrix and $T^{N_fN_f}$ and also $B$ into light flavored ${\tilde B}$ and heavy flavored parts. The off-diagonal elements of $T$ and the heavy flavored $B$ vanish at the minimum, where $T^{N_fN_f}$ = $S/m$ is derived.  Inserting this relation into Eq.(\ref{Eq:MassiveWeff}),  we finally obtain
\begin{equation}\label{Eq:MassiveDecoupledWeff}
W_{\rm eff}=S 
\left\{ 
\ln\left[
\frac{
	S^{N_c-N_f-1}{\rm det}({\tilde T}) f({\tilde Z})
}
{
	{\tilde \Lambda}^{3N_c-N_f-5}
} 
\right] 
+N_f-N_c+1\right\},
\end{equation}
where ${\tilde Z}$ = ${\tilde B}{\tilde T}^{N_f-N_c-1}{\tilde B}$/${\rm det}({\tilde T})$ from $Z$ = ${\tilde B}T^{N_fN_f}{\tilde T}^{N_f-N_c-1}{\tilde B}/T^{N_fN_f}{\rm det}({\tilde T})$ and ${\tilde \Lambda}^{3N_c-N_f-5}$ = $m\Lambda^{3N_c-N_f-6}$.  Thus, we are left with Eq.(\ref{Eq:Weff}) with $N_f-1$ massless flavors.  This decoupling is successively applied to the $SO(N_c)$ theory until $N_f$ is reduced to $N_c$.

In the massless $SO(N_c)$ theory, it should be noted that if $S$ is integrated out\cite{Sout}, one reaches the following superpotential for the massless $SO(N_c)$ theory:
\begin{equation}\label{Eq:W_ADS}
W_{\rm eff}^\prime=\left( N_f-N_c+2\right) \left[ 
\frac{
	{\rm det}\left(T\right) f(Z)
}
{
	\Lambda^{3N_c-N_f-6}
} 
\right]^{1/(N_f-N_c+2)}. 
\end{equation}
Since $W_{\rm eff}^\prime$ vanishes in the classical limit, where the constraint of $BT^{N_f-N_c}B$ = ${\rm det}\left(T\right)$, namely $Z$ = 1, is satisfied, we find that $f(1)$ = 0.  The simplest form of $f(Z)$ can be given by
\begin{equation}\label{Eq:f_Z}
f(Z)=1-Z, 
\end{equation}
which gives
\begin{equation}\label{Eq:W_OURS}
W_{\rm eff}^\prime=\left( N_f-N_c+2\right) \left[ 
\frac{
	{\rm det}\left(T\right)- BT^{N_f-N_c}B
}
{
	\Lambda^{3N_c-N_f-6}
} 
\right]^{1/(N_f-N_c+2)}. 
\end{equation}

The massive $SO(N_c)$ theory can also be analyzed by considering the instanton contributions from the gauginos and $N_f$ quarks in the massless $SO(N_c)$ theory. The instanton contributions can be expressed as 
\begin{equation}
\label{Eq:Instanton}
(\lambda\lambda)^{N_c-2}{\rm det}(\psi^i\psi^j), 
\end{equation}
where $\psi$ ($\lambda$) is a spinor component of $Q$ ($S$).  In the case with $N_f-N_c+2$ massive quarks labeled by $i$=$N_c-1$ $\sim$ $N_f$, this instanton amplitude is further transformed into\cite{MassiveInstanton}
\begin{equation}
\label{Eq:MassiveInstanton}
\prod_{a=1}^{N_c-2}\pi_a = c\Lambda^{2N_c-4}\prod_{i=Nc-1}^{N_f}m_i/\Lambda, 
\end{equation}
where $\pi_i$ represents the scalar component of $T^{ii}$ and $c$ is a non-vanishing coefficient.  On the other hand, our superpotential yields the conditions of $\partial W_{\rm eff}/\partial\pi_\lambda$ = 0 and $\partial W_{\rm eff}/\partial\pi_i$ = 0 for $i$ = $N_c-1$ $\sim$ $N_f$ and gives $\pi_\lambda/\pi_i$ = $m_i$ ($i$ = $N_c$ $\sim$ $N_f$) and $(1-\alpha)\pi_\lambda/\pi_i$ = $m_i$ ($i$ = $N_c-2$ and $N_c-1$) known as the Konishi anomaly relation\cite{KonishiAnomaly}, where $\alpha$ = $zf^\prime (z)/f(z)$ for $z$ = $\langle 0 \vert Z \vert 0 \rangle$.  We find that 
\begin{equation}
\label{Eq:MassiveFunc}
(1-\alpha)^2f(z) = \prod_{a=1}^{N_c-2}\Lambda^2/\pi_a\prod_{i=N_c-1}^{N_f}m_i/\Lambda.
\end{equation}
These two relations of Eqs.(\ref{Eq:MassiveFunc}) and (\ref{Eq:MassiveInstanton}) show that the mass dependence in $f(z)$ is completely cancelled.  The absence of this mass dependence is a clear indication of the consistency of our superpotential of Eq.(\ref{Eq:Weff}). Since $(1-\alpha)^2f(z)$ = 1/$c$ ($\neq$0), we obtain that $f(z)\neq$0 or equivalently $z\neq 1$ for Eq.(\ref{Eq:f_Z}).  Therefore, the classical constraint, corresponding to $z$ = 1, should be modified for this massive $SO(N_c)$ theory.


Now, we discuss the vacuum structure in the massless $SO(N_c)$ theory.  It is obvious that our superpotential allows all vacuum expectation values (VEV's) to vanish\cite{SO}.  In this case, since the anomaly-matching is not satisfied by original massless composite fields, the $SO(N_c)$ theory should employ magnetic degree's of freedom called magnetic quarks\cite{SO}.  However, we will advocate the possible existence of another vacuum with non-vanishing VEV's determined by the superpotential of Eq.(\ref{Eq:W_OURS}), which yields dynamical breakdown of the flavor symmetry of $SU(N_f) \times U(1)_R$ compatible with the anomaly-matching property. Both vacua become the correct vacua with the same vanishing vacuum energy since they are supported by appropriate effective superpotentials described by either magnetic quarks or by composites made of the original quarks.

To find the spontaneous symmetry-breaking solution, we rely upon on the plausible theoretical expectation that the SUSY theory has a smooth limit to the SUSY-preserving phase from the SUSY-breaking phase.  In order to see solely the SUSY breaking effect, we adopt the simplest term that is invariant under the whole flavor symmetries, which is given by the following mass term, ${\cal L}_{mass}$, for the scalar quarks, $\phi_A^i$:
\begin{equation}\label{Eq:SUSYBreakingTerm}
-{\cal L}_{mass}=\mu^2\sum_{i,A}\vert \phi_A^i \vert^2. 
\end{equation}
 Together with the potential term arising from $W_{\rm eff}$, we find that
\begin{eqnarray}
& & V_{\rm eff}
=
G_T \bigg( 
   \sum_{i=1}^{N_f}  |W_{{\rm eff};i}|^2         
\bigg)
+ 
G_B|W_{{\rm eff};B}|^2
+
G_S |W_{{\rm eff};\lambda} |^2
+
V_{\rm soft}, 
\label{Eq:Veff} \\
& & V_{\rm soft}
=\mu^2\Lambda^{-2}\sum_{i=1}^{N_f}|\pi_i|^2
+
\Lambda^{-2(N_c-1)}\mu^2|\pi_B|^2
\label{Eq:Vsoft}
\end{eqnarray}
with the definition of $W_{{\rm eff};i}$ $\equiv$  $\partial W_{\rm eff}/\partial \pi_i$, etc., where $\pi_\lambda$ and $\pi_B$, respectively,  represent the scalar components of $S$ and $B^{[12\cdots N_c]}$.
\footnote{Other fields such as scalar components of $T^{ij}$ with $i\neq j$ can be set to vanish at the minimum of $V_{\rm eff}$; therefore, we omit these terms.}
 The coefficient $G_T$ comes from the K$\ddot{\rm a}$hlar potential, $K$, which is assumed to be diagonal, $\partial^2 K/\partial T^{ik\ast}\partial T^{j\ell}$ = $\delta_{ij}\delta_{k\ell}G_T^{-1}$ with $G_T$ = $G_T(T^{\dagger}T)$, and similarly for $G_B$ = $G_B(B^{\dagger}B)$ and $G_S$ = $G_S(S^{\dagger}S)$.  Since we are interested in the SUSY-breaking phase in the vicinity of the SUSY-preserving phase, the leading terms of $\mu^2$ are sufficient to control the SUSY breaking effects.  Namely, we assume that $\mu^2/\Lambda^2\ll 1$.

Since the original flavor symmetries do not provide the consistent anomaly-matching property, the $SO(N_c)$ dynamics requires that some of the $\pi$ acquire non-vanishing VEV's so that it triggers dynamical symmetry breaking that supplies the Nambu-Goldstone bosons to produce appropriate anomalies.  In the SUSY limit, this breaking simultaneously supplies massless quasi Nambu-Goldstone fermions that also produce appropriate anomalies.  If all anomalies are saturated by these Nambu-Goldstone superfields, flavor symmetries cease to undergo further dynamical breakdown.  Therefore, the $SO(N_c)$ dynamics at least allows one of the $\pi_i$ ($i$=1 $\sim$ $N_f$) to develop a VEV and let this be labeled by $i$ = $1$: $|\pi_1|$ = $\Lambda_T^2$ $\sim$ $\Lambda^2$. This VEV is determined by solving  $\partial V_{\rm eff}/\partial\pi_i$ = 0 that yields
\begin{equation}\label{Eq:Tab}
G_TW_{{\rm eff};a}^\ast
\frac{\pi_\lambda}{\pi_a} \left( 1-\alpha \right) 
 =  G_SW_{{\rm eff};\lambda}^\ast \left( 1-\alpha \right)+\beta X_B +M^2\big| \frac{\pi_a}{\Lambda}\big|^2, 
\end{equation}
for $a$=1${\sim}N_c$,
\footnote{If we choose $i$=$N_c+1\sim N_f$, which are not the indices used in $B^{[12\cdots N_c]}$, we cannot obtain the consistent vacuum configuration.}
 where $\beta$ = $z\alpha^\prime$
and
\begin{eqnarray}
& & 
M^2 
= 
\mu^2 +G_T^\prime\Lambda^2\sum_{i=1}^{N_f} \big| W_{{\rm eff};i}\big|^2 
\label{Eq:ScalarMass}, \quad
X_B 
=
G_T\sum_{a=1}^{N_c}W_{{\rm eff};a}^\ast\frac{\pi_\lambda}{\pi_a}
-
G_BW_{{\rm eff};B}^\ast\frac{\pi_\lambda}{\pi_B}. 
\label{Eq:X_B}
\end{eqnarray}
The SUSY breaking effect is specified by $\mu^2\vert \pi_1 \vert^2$ in Eq.(\ref{Eq:Tab}) through $M^2$ because of $\pi_1$ $\neq$ 0.  This effect is also contained in $W_{{\rm eff};\lambda}$ and $X_B$. From Eq.(\ref{Eq:Tab}) with $\partial W_{{\rm eff};a} = (1-\alpha)\pi_\lambda/\pi_a$ ($a=1\sim N_c$), we find that  
\begin{equation}
\bigg| \frac{\pi_a}{\pi_1} \bigg|^2 = 1 + 
\frac{(M^2/\Lambda^2)(\big| \pi_1\big|^2-\big| \pi_a\big|^2)} {G_SW_{{\rm eff};\lambda}^\ast \left( 1-\alpha \right)+\beta X_B +
(M^2/\Lambda^2)\big| \pi_a\big|^2}.
\label{Eq:VEVS}
\end{equation}
It is obvious that $\pi_{a\neq 1}$ = 0 cannot satisfy Eq.(\ref{Eq:VEVS}).  In fact, $\pi_{a\neq 1}$ = $\pi_1$ is a solution to this problem, leading to $|\pi_a|$ = $|\pi_1|$ (= $\Lambda_T^2$). As a whole, all the $|\pi_a|$ for $a$=$1\sim N_c$ dynamically acquire the same VEV once one of these $|\pi_a|$ receives a VEV. 

Let us estimate each VEV for composite superfields. Since the classical constraint of $f(z)$ = 0 is to be proved to receive no modification at the SUSY minimum, this breaking effect can arise as tiny deviation of $f(z)$ from 0 in the SUSY-breaking vacuum, which is denoted by $\xi$ $\equiv$ $1 - z$ ($\ll$ 1).  After a little calculus, we find
\begin{equation}
\label{Eq:SUSYbreakingVEV}
\big|\pi_{i=1 \sim N_c} \big|  = \Lambda_T^2, \ \
\big|\pi_{i=N_c+1 \sim N_f} \big|  = \xi\big|\pi_{i=1 \sim N_c} \big|, \ \
|\pi_B| \sim \Lambda_T^{N_c},  \ \
\big|\pi_\lambda\big| \sim \Lambda^3\xi^{\frac{N_f-N_c+1}{N_f-N_c+2}}, 
\end{equation}
in the leading order of $\xi$, whose $\mu$-dependence can be fixed by solving the set of equations of $\partial V_{\rm eff}/\partial\pi_{i,B,\lambda}$ = 0.  In the softly broken $SO(N_c)$ theory, our superpotential indicates the breakdown of all chiral symmetries.

We expect that this phenomenon persists to occur in the SUSY limit of $\mu\rightarrow 0$ or equivalently $\xi\rightarrow 0$.  At the SUSY minimum with the suggested vacuum of $|\pi_{a=1\sim N_c}|$ = $\Lambda_T^2$,  we find the classical constraint of $f(z)$ = 0, which is derived by using $W_{{\rm eff};\lambda}$ = 0 and by noticing that $\pi_\lambda/\pi_{i=N_c+1 \sim N_f}$ = 0 from $W_{{\rm eff};i}$ = 0. In the SUSY limit, $\pi_{i=N_c+1 \sim N_f}$ vanish to recover chiral $SU(N_f-N_c)$ symmetry and $\pi_\lambda$ vanishes to recover chiral $U(1)$ symmetry.  The symmetry breaking is thus described by 
\begin{eqnarray}
\label{Eq:ChiralSymmetryBreaking}
& &
SU(N_f) \times U(1)_R \rightarrow SO(N_c) \times SU(N_f-N_c) \times U(1)^\prime_R, 
\end{eqnarray}
where $U(1)^\prime_R$ is associated with the number of $SU(N_c)$-adjoint and $SU(N_c)$-singlet fermions and with the number of the ($N_f-N_c$)-plet scalars of $SU(N_f-N_c)$. 

This dynamical breaking can really persist in the SUSY limit only if the anomaly-matching property is respected by composite superfields. The anomaly-matching property is easily seen by the use of the complementarity\cite{tHooft,Susskind}.  This dynamical breakdown of $SU(N_f) \times U(1)_R$ can also be realized by the corresponding Higgs phase defined by $\langle 0| \phi_A^a|0\rangle$ = $\delta_A^a\Lambda_T$ for $a,A$ = 1 $\sim$ $N_c$. The anomaly-matching is trivially satisfied in the Higgs phase, whose spectrum is found to be precisely equal to that of the massless composite Nambu-Goldstone superfields represented by $T^{ab}$ with Tr($T^{ab}$) = 0, $T^{ia}$ and a linear combination of Tr($T^{ab}$) and $B^{[12\cdots N_c]}$ ($a,b$=$1\sim N_c$, $i$=$N_c+1\sim N_f$).  No other composite fields such as those including gauge chiral superfields are not present in the low-energy spectrum.  Our superpotential, thus, assures that the anomaly-matching is a dynamical consequence.

Summarizing our discussions, we have examined the dynamics of the $SO(N_c)$ theory with $N_c\leq N_f$ $\left( < 3(N_c-2)\right)$ based on the effective superpotential:
\begin{equation}
\label{Eq:W_eff_summary}
W_{\rm eff}=S 
\left\{ 
\ln\left[
\frac{
	S^{N_c-N_f-2}\left(
	{\rm det}\left(T\right)- BT^{N_f-N_c}B\right)
}
{
	\Lambda^{3N_c-N_f-6}
} 
\right] 
+N_f-N_c+2\right\}+\sum_{i=1}^{N_f}m_iT^{ii}.
\end{equation}
This superpotential is equivalent to the following superpotential:
\begin{equation}
W_{\rm eff}^\prime=\left( N_f-N_c+2\right) \left[ 
\frac{
	{\rm det}\left(T\right)- BT^{N_f-N_c}B
}
{
	\Lambda^{3N_c-N_f-6}
} 
\right]^{1/(N_f-N_c+2)}+\sum_{i=1}^{N_f}m_iT^{ii}.
\label{Eq:WEFFADS}
\end{equation}
We have found that our superpotential respects 
\begin{enumerate}
\item holomorphic decoupling property, 
\item correct vacuum structure with instanton physics when $N_c-2$ quarks become massless and others remain massive, 
\item dynamical breakdown of the flavor $SU(N_c)$ symmetry described by 
\begin{eqnarray}
\label{Eq:SymmetryBreakingSummary}
& &SU(N_f) \times U(1)_R \rightarrow SO(N_c) \times SU(N_f-N_c) \times U(1)^\prime_R, 
\end{eqnarray}
in the massless $SO(N_c)$ theory and 
\item consistent anomaly-matching property due to the emergence of the Nambu-Goldstone superfields. 
\end{enumerate}
The explicit VEV's for composite superfields induced by our flavor-blind soft SUSY-breaking are calculated to be:
\begin{eqnarray}
& 
\big|\langle 0 |T^{ij}|0\rangle\big| = 
\left\{ \begin{array}{l}
	 \Lambda_T^2\delta^{ij}~~~(i=1 \sim N_c) \\
	\xi\Lambda_T^2\delta^{ij}~~(i=N_c+1 \sim N_f) \\
\end{array} \right.,
\nonumber \\
& \big|\langle 0 |B^{[12\cdots N_c]}|0\rangle\big| \sim \Lambda_T^{N_c}, \ \ \
\big|\langle 0 |S|0\rangle\big| \sim \Lambda_T^3\xi^{\frac{N_f-N_c+1}{N_f-N_c+2}},
\end{eqnarray}
where $\xi$ $\rightarrow$ 0 gives the SUSY limit.  The estimation clearly shows that the onset of the flavor-blind soft SUSY-breaking completely breaks the residual chiral symmetries of $SU(N_f-N_c)$ and $U(1)^\prime_R$.  The classical constraint of ${\rm det}\left(T\right)$=$BT^{N_f-N_c}B$ is modified into
\begin{equation}
{\rm det}\left(T\right) - BT^{N_f-N_c}B = \xi {\rm det}\left(T\right),
\end{equation}
which measures the parameter $\xi$.

Since the dynamical flavor symmetry breaking can be consistently described by $W_{\rm eff}^\prime$ of Eq.(\ref{Eq:WEFFADS}), we speculate that the $N$=1 physics controlled by $W_{\rm eff}^\prime$ reflects the remnant of the $N$=2 physics near the Chebyshev point, where the dynamical flavor symmetry breaking is a must. We also hope that the advent of the powerful theoretical tools to implement chiral fermions in lattice formulation\cite{CF} of the $SO(N_c)$ dynamics and of powerful computers to perform lattice simulation reveals the real physics of the $SO(N_c)$ theory.

\begin{center}
{\bf ACKNOWLEDGMENTS}
\end{center}
This work is supported by the Grants-in-Aid for Scientific Research on Priority Areas B (No 13135219) from the Ministry of Education, Culture, Sports, Science, and Technology, Japan.

\end{document}